# Plasmonic Instabilities in Two-Dimensional Electron Channels of Variable Width

G. R. Aizin[1†], J. Mikalopas[1], and M. Shur[2,3*]

[1] *Kingsborough College, The City University of New York, Brooklyn, New York 11235, USA*

[2] *Rensselaer Polytechnic Institute, Troy, New York 12180, USA*

[3] *Electronics of the Future, Inc., Vienna VA 22181, USA*

**Abstract**

Understanding of fundamental physics of plasmonic instabilities is the key issue for the design of a new generation of compact terahertz electronic sources required for numerous THz applications. Variable width plasmonic devices have emerged as potential candidates for such an application. The analysis of the variable width plasmonic devices presented in this paper shows that these structures enable both the Dyakonov-Shur instability (when the electron drift velocity everywhere in the device remains smaller than the plasma velocity) and the "plasmonic boom" instability that requires drift velocity exceeding the plasma velocity in some of the device sections. For symmetrical structures, the driving current could be provided by an RF signal leading to RF to Terahertz (THz) and THz to RF frequency conversion using the source and drain antennas and reducing losses associated with ohmic contacts. We show that narrow regions protruding from the channel ("plasmonic stubs") could control and optimize boundary conditions at the contacts and/or at the interfaces between different device sections. These sections could be combined into plasmonic crystals yielding enhanced power and a better impedance matching. The mathematics of the problems is treated using the transmission line analogy. We show that the combination of the stubs and the variable width channels is required for the instability rise in an optimized plasmonic crystal. Our estimates show that THz plasmonic crystal oscillators could operate at room temperature.

[†]gaizin@kbcc.cuny.edu

[*]shurm@rpi.edu

# I. INTRODUCTION

Short channel field effect transistors (FETs) supporting decaying or resonant plasma waves have promise of invigorating the THz electronics by providing efficient THz and sub-THz detectors and sources [1,2]. Such transistors are sometimes called "TeraFETs". TeraFET detectors using silicon complementary metal-oxide semiconductor (CMOS) FETs [3], InGaAs High Electron Mobility Transistors (HEMTs) [4], AlGaN/GaN HEMTs [5], and graphene [6] operated in the 0.1 THz to 22 THz range [7] with Noise Equivalent Power (NEP) as low as 0.5 pW/Hz$^{1/2}$ [8]. p-diamond [9] might have advantages for implementing TeraFETs operating in 240 GHz to 320 GHz range for beyond 5G WIFI operation. TeraFET sources use various plasma instabilities in the two-dimensional (2D) electron channel of the FET to generate electromagnetic radiation in the THz range. However, the THz emission associated with the Dyakonov-Shur (DS) instability [10] has yielded very low powers and the THz emission based on the proposed "plasmonic boom" mechanism [11,12] has only been observed very recently for the first time [13]. Furthermore, the estimates of power budget for beyond 5G WI FI applications [14] reveal that further improvements of NEP down to at 0.1 pW/Hz$^{1/2}$ are needed for this potential "killer application" of the sub-THz technology.

The ideas to improve both detection and emission of the sub-THz and THz radiation by TeraFETs focused on several approaches: moving from a single TeraFET to a "plasmonic crystal" [15,16], using the TeraFET asymmetric multi-gate structure [17,18] and, more recently, using "plasmonic stubs" - narrow regions protruding from the channel and having tunable electrical parameters [19,20]. The stubs allow for an optimization and adjustment of the boundary conditions at the contacts and/or at the interfaces between different plasmonic cavities [19] providing more favorable conditions for the DS instability. As shown in Ref. [19], the stubs could also adjust and tune the plasma velocity. This capability should make it easier to reach the plasmonic boom conditions. In Ref. [20], it was shown that the stubs enable the DS instability for both directions of the driving source-to-drain current. Hence these structures could be driven by an RF signal to excite the plasmonic instability and generate the THz radiation. The device channels with the stubs must be narrow so that the channel width and the stub protrusion remain smaller than or comparable to the electron mean free path in order to achieve the ballistic transport condition. The stub TeraFET is shown schematically in Fig. 1b among other suggested TeraFET configurations.

We define the stub as a protrusion with the size in the source-drain direction being much smaller than the channel plasmonic wavelength. [19]. In this paper, we analyze the TeraFET plasmonic devices with a varying width of the channel where the lengths of the sections of different width (Fig. 1c) are comparable to the plasmonic wavelength in the channel (variable width FETs). The varying width affects the current distribution in the channel as explained in Section 2. We consider gated TeraFETs with the gate-to-channel separation much smaller than the gate length and width in any section of the device and use the plane capacitor model for description of the quasi-static electric interaction between the gate and the channel

Our analysis demonstrates that these structures enable both the DS instability (when the electron drift velocity everywhere in the device remains smaller than the plasma velocity) and the "plasmonic boom" instability that requires drift velocity exceeding the plasma velocity in some of the device sections. The stub and the variable width FETs could be combined to form one-dimensional plasmonic crystals (Fig. 1d) and two-dimensional patterns (Fig. 1e). We show that in these structures, the driving current could also be provided by the RF signal leading to RF to THz and THz to RF frequency conversion. Such conversion using the source and drain antennas reduces losses associated with ohmic contacts as schematically illustrated in Fig 1f.

The rest of this paper is organized as follows. In Sec. II, we develop the general theory of the variable width plasmonic devices with comparable lengths of the sections of different widths. Our model is based on the transmission line (TL) approach and describes the structures with finite number of sections with different width as well as the plasmonic crystals based on these structures. In Sec. III, we apply the developed model for the analysis of the plasma instability in the TeraFET with three sections of variable width. The advantage of these structures is that they enable both the DS instability and the "plasmonic boom" instability. In Sec. IV, we use the developed approach to analyze performance of the plasmonic crystals with periodically changing channel width. In Sec. V, we present numerical estimates for different materials used in the TeraFET structures. These estimates yield the parameters ranges for the experimental observation of the predicted effects at cryogenic and room temperatures. Our conclusions and a brief summary are presented in Sec. V. Some additional analysis is contained in the Appendix.

## II. BASIC EQUATIONS

We consider plasma oscillations in the 2D electron channel of variable width in the FET biased by a DC current. The general FET geometry is shown in Fig. 1c. The FET channel consists of several segments with different constant widths. We assume that the widths of the individual segments are much smaller than their lengths and consider plasma waves propagating along the channel ($x$-axis) between the source and the drain. In the hydrodynamic approximation, the plasma waves in the 2D electron layer ($z=0$) are described by the Euler equation and the equation of continuity as

$$\frac{\partial v}{\partial t} + v\frac{\partial v}{\partial x} = \frac{e}{m^*}\frac{\partial \varphi}{\partial x}$$

(1)

$$\frac{\partial n}{\partial t} + \frac{\partial (nv)}{\partial x} = 0,$$

where $n(x,t)$ and $v(x,t)$ are the local electron density and the velocity in the plasma wave, $-e$ is the electron charge, $m^*$ is the electron effective mass, and $\varphi(x, z = 0, t)$ is electric potential in the 2D layer. This hydrodynamic approach is justified when the electron-electron scattering length $l_{ee}$ is the shortest characteristic length in the system. In particular, the value of $l_{ee}$ should be smaller than the size of the transient regions between the segments of different width in the FET channel. This condition makes hydrodynamic model applicable for description of the plasma waves in the entire 2D channel of the variable width. We also assume ballistic electron transport with respect to the collisions with phonons and impurities and neglect the collision term in the Euler equation. This approximation is justified if $\omega_p \tau \gg 1$ where $\omega_p$ is the plasma frequency and $\tau$ is the collision time. The pressure gradient term is also omitted in the Euler equation because in the gated 2D channel this term is small in compared with the field term [12].

To obtain plasma waves, we linearize the hydrodynamic equations (1) with respect to the small fluctuations of electron density $\delta n(x,t)$ and velocity $\delta v(x,t)$ assuming that $n = n_0 + \delta n$ and $v =$ $v_0 + \delta v$ where $n_0$ and $v_0$ are the equilibrium electron density and the constant electron drift velocity respectively. In the gradual channel approximation, the fluctuations of the electron density $\delta n$ and electric potential $\delta\varphi$ in the gated 2D channel are connected as $-e\delta n = C\delta\varphi$ where $C = \varepsilon\varepsilon_0/d$ is the gate-to-channel capacitance per unit area, $d$ is the gate-to-channel separation and $\varepsilon$ is the dielectric constant of the barrier layer between the channel and the gate. In this model, general solution of the linearized Eq. (1) describing the plasma wave with frequency $\omega$ and wave vector $q$ ($\delta n, \delta v \propto exp(-iqx + i\omega t)$) propagating in the 2D channel is [12]

$$I_\omega = I_1 e^{-iq_1 x} + I_2 e^{-iq_2 x}$$

(2)

$$V_\omega = \frac{1}{CW}\left(\frac{I_1}{v_0+v_p}e^{-iq_1 x} + \frac{I_2}{v_0-v_p}e^{-iq_2 x}\right),$$

where $I_\omega = W\delta j_\omega = -eW(v_0\delta n_\omega + n_0\delta v_\omega)$ is the total plasmonic current in the 2D channel of width $W$, $V_\omega \equiv \delta\varphi_\omega(x)$ is the voltage distribution in the plasma wave, $q_{1,2} = \frac{\omega}{v_0 \pm v_p}$ are the wave vectors of the plasmons propagating in the direction of the constant electron drift, $q_1$, and in the opposite direction, $q_2$, $v_p = \sqrt{e^2 n_0/m^* C}$ is the plasma velocity of the gated plasmon at $v_0 = 0$. Coefficients $I_1$ and $I_2$ depend on $v_0$ and $v_p$ and should be determined from the boundary conditions.

It has been demonstrated in numerous publications [21 - 23] that the linearized hydrodynamic equations (1) describing the plasma waves in the gated 2D channel at $v_0 = 0$ have the form of the telegrapher's equations for the TL with the distributed inductance $\mathcal{L} = m^*/e^2 n_0 W$, distributed capacitance $CW$, and, in the presence of disorder, distributed resistance $\mathcal{R} = \mathcal{L}/\tau$ so that the gated electron channel can be considered as a plasmonic waveguide [24 - 26]. In Ref. [12], it has been shown that the TL analogy still stands in the presence of a constant electron drift with velocity $v_0$ in the channel if the electric voltage $V_\omega$ in the telegrapher's equations is replaced with the effective voltage $V_\omega^{eff} = V_\omega + V_\omega^{kin}$ where $V_\omega^{kin} = -m^* v_0 \delta v/e$ is the so called kinetic voltage [27] related to the kinetic power carried by the oscillating electrons in the drifting plasma wave. It follows from Eq. (2) that [12]

$$V_\omega^{eff} = V_\omega + V_\omega^{kin} = (1 - M^2)V_\omega + Z_0 M I_\omega, \tag{3}$$

where $M = \frac{v_0}{v_p}$ is the Mach number, and $Z_0 = \sqrt{\mathcal{L}/CW} = 1/CW v_p$ is the characteristic impedance of the plasmonic TL. In the hydrodynamic model, the continuity of $V_\omega^{eff}$ and $I_\omega$ corresponds to the conservation of energy and of the number of particles in the plasma wave maintained in the entire 2D channel.

Using Eqs. (2) and (3) one can relate the values of $V_\omega^{eff}$ and $I_\omega$ at the opposite ends of the plasmonic waveguide representing the 2D electron channel. For a DC biased 2D channel of length $l$ and constant width $W$ positioned between the source (s) and the drain (d) we obtain [12]

$$\begin{pmatrix} V_{\omega,s}^{eff} \\ I_{\omega,s} \end{pmatrix} = \hat{t} \begin{pmatrix} V_{\omega,d}^{eff} \\ I_{\omega,d} \end{pmatrix}, \tag{4}$$

where

$$\hat{t} = e^{-i\frac{v_0}{v_p}\Theta} \begin{pmatrix} \cos\Theta & \frac{i}{WCv_p}\sin\Theta \\ iWCv_p\sin\Theta & \cos\Theta \end{pmatrix}, \quad \Theta = \frac{\omega \ell v_p}{v_p^2 - v_0^2}. \tag{5}$$

Eq. (4) can also be applied for description of the FET with the 2D electron channel consisting of several segments of different length and width provided that the transient regions between the segments meet restrictions imposed by the hydrodynamic model as discussed at the beginning of this section. For the segmented channel, the matrix $\hat{t}$ in Eq. (4) should be replaced with the matrix

$$\hat{T} = \hat{t}_1 \hat{t}_2 ..., \tag{6}$$

where matrices $\hat{t}_j$ are described by matrix $\hat{t}$ in Eq. (5) written for each individual segment $j$, $j = 1,2,...$.

The plasmon dispersion equation in the segmented plasmonic cavity characterized by the transfer matrix $\hat{T}$ and bounded by the source and the drain contacts can be found by adding the boundary conditions to Eqs. (4) - (6) at the contacts. These boundary conditions depend on the impedances between the source and the gate, $Z_{gs}$, and the drain and the gate, $Z_{gd}$, and can be written as [20]

$$V_{\omega,s} = -Z_{gs} I_{\omega,s}$$

$$V_{\omega,d} = Z_{gd} I_{\omega,d} \tag{7}$$

For the sake of simplicity, we consider symmetric structures with $Z_{gs} = Z_{gd} \equiv Z_g$ and the width $W$ of the 2D channel being the same near the source and the drain contacts, though the results can be easily generalized to the arbitrary asymmetric boundaries. Combining Eqs. (3) - (7) we obtain the following plasmon dispersion equation in the source-drain cavity

$$(M^2 - 1)^2 T_{21} Z_g^2 - (M^2 - 1)(T_{11} + T_{22})Z_g + Z_0 M(T_{11} - T_{22}) - Z_0^2 M^2 T_{21} + T_{12} = 0 \tag{8}$$

In the next Section we present solution of Eq. (8) and the analysis of the obtained results.

If a large number of identical plasmonic cavities described above are connected together the device behaves as a plasmonic crystal with repeated segmented elementary cells each characterized by the transfer matrix $\hat{T}$. The 2x2 matrix $\hat{T}$ connects the values of $V_\omega^{eff}$ and $I_\omega$ at the opposite ends of the cell of length $L$. Using the Bloch condition $[V_\omega^{eff}(x+L), I_\omega(x+L)] = e^{ikL}[V_\omega^{eff}(x), I_\omega(x)]$ where $k$ is the Bloch wave vector defined in the region $-\pi/L \leq k \leq \pi/L$, we obtain the following dispersion equation for the plasmonic crystal

$$det\hat{T} - e^{ikL} Tr\,\hat{T} + e^{2ikL} = 0 \tag{9}$$

This equation (identical to the one used in the photonics crystal theory [28]) is used in Section IV to derive plasmon dispersion in the plasmonic crystal TeraFET.

### III.   PLASMA INSTABILITIES IN TeraFET OF VARIABLE WIDTH

To explore the plasma instabilities in the FET of variable width we applied Eq. (8) to the symmetric FET structure shown in Fig. 2a. The TL model for this FET structure is shown in Fig. 2b.

This FET structure consists of three individual segments of equal length $L$ but the width of the central segment $W_1$ is larger than the width $W$ of the two identical segments adjacent to the source and drain contacts. We also assumed that $Z_{gs} = Z_{gd} = 0$ so that an ac current at the drain and source contacts is short circuited to the gate. In this case, Eqs. (5), (6) and (8) yield the following dispersion equation

$$\frac{2\gamma}{\gamma^2+1}\sin\frac{2\omega L}{v_p(1-M^2)}\cos\frac{\omega L}{v_p(1-\gamma^2 M^2)} + \sin\frac{\omega L}{v_p(1-\gamma^2 M^2)}\left(\cos\frac{2\omega L}{v_p(1-M^2)} - \frac{1-\gamma^2}{1+\gamma^2}\frac{1+M^2}{1-M^2}\right) = 0 \tag{10}$$

where $\gamma = \frac{W}{W_1} < 1$, and the Mach number is defined with respect to the drift velocity $v_0$ in the narrow segments adjacent to the FET contacts, see Fig. 2.

First, we consider Eq. (10) when $\gamma = 0$. In this limit, we obtain two sets of solutions

$$\omega_n^{(1)} = \frac{\pi v_p}{L}n, \quad n = 1, 2, \ldots \tag{11}$$

and

$$Re\,\omega_n^{(2)} = \frac{\pi v_p |1 - M^2|}{2L}n, \quad n = \begin{cases} 2, 4, \ldots & |M| < 1 \\ 1, 3, \ldots & |M| > 1 \end{cases}$$

$$Im\,\omega_n^{(2)} = \pm\frac{v_p(1-M^2)}{2L}\ln\left|\frac{1+M}{1-M}\right| \tag{12}$$

Fig. 3a and 3b show these solutions as functions of $|M|$, $0 < |M| < 2$. The first solution, $\omega_n^{(1)}$, corresponds to the energy levels of plasmons confined in the central wide region of the FET structure, see Fig. 2. The limit $\gamma = 0$ implies the infinitely wide central cavity, where the drift velocity effectively reduces to zero and the plasmon energy spectrum in Eq. (11) does not depend on $M$ as shown in Fig. 3a. This spectrum describes the plasmon in the cavity with symmetric boundaries.

The second solution, $\omega_n^{(2)}$, describes the energy spectrum of the plasmons confined in the side regions of the FET channel adjacent to the source (the source cavity) or to the drain (the drain cavity). These cavities have asymmetrical boundaries leading to the Dyakonov-Shur plasma instability, and the plasma spectrum in Eq. (12) is similar to the spectrum first derived by Dyakonov and Shur in Ref. [1]. However, there is a very important difference. In the limit of $\gamma = 0$, the oscillating plasma current density at the boundary between the source (the drain) cavity and the central cavity is zero at the central cavity side because the current spreads out into the central section of the infinite width. As the total current is preserved, the spatial derivative of the total current at this boundary at M=0 is zero, and, therefore, the boundary condition for the current density at the source (drain) side corresponds to the antinode of the standing wave. Hence, at $M = 0$, the standing plasma waves confined in these cavities have the integer number of half wavelengths being equal to the length of the cavity. Inset in Fig. 3b showing the spatial distribution of the normalized plasmonic current in the source cavity for several values of $M > 0$ clearly illustrates this point explaining the difference in the wave numbers of the plasma modes: $n = 1,3,5, ...$ for the Dyakonov-Shur instability [1] and $n = 2,4,6, ...$ in Eq. (12) for $|M| < 1$ and vice versa for $|M| > 1$.

The instability occurs for the modes with $Im\omega_n^{(2)} < 0$. As shown in Ref [1] the plasmonic instability develops in the DC biased asymmetric cavity if the current flows from the low impedance cavity edge to the high impedance one at $0 < M < 1$ and in the opposite direction if $M < -1$. In the variable width FET (Fig. 2), the instability develops for any direction of the DC current and at any value of $M$ as it follows from Eq. (12). For $0 < M < 1$ and $M < -1$, this instability is supported by the source cavity and for $-1 < M < 0$ and $M > 1$, it is supported by the drain cavity. This conclusion follows from the obvious fact that the source and the drain cavities have reflection symmetry. In Fig. 3, we plotted $Im\omega_n^{(2)}$ for both stable ($Im\omega_n^{(2)} > 0$) and unstable ($Im\omega_n^{(2)} < 0$) modes.

For a more relevant case of $\gamma \neq 0$, the plasmons in the individual cavities interact. In particular, at $M = 0$, the plasmon energy in the central cavity is still determined by Eq. (11) but the plasmon energy levels in the source and the drain cavities split.

$$\omega_n^{(2)} = \frac{\pi v_p}{L}n \pm \frac{v_p}{L}arccos\frac{1}{1+\gamma}, n = 1,2, ... \qquad (13)$$

At finite values of $M$ and $\gamma \neq 0$, the plasmon energy levels in the central cavity depend on $M$:

$$\omega_n^{(1)} = \frac{\pi v_p}{L}|1 - \gamma^2 M^2|n, n = 1,2, .... \qquad (11a)$$

Eq. (11a) represents the plasmon spectrum of the drifting plasmon in a symmetric cavity [12]. At finite $\gamma$, interaction between the plasma modes in different cavities results in multiple anticrossings and merging of the plasmon dispersion curves in the central cavity with the ones in the source/drain cavities. This dependence modifies both the real and imaginary components of the plasma frequencies as discussed below.

Eq. (10) was solved numerically for finite values of $M$ and $\gamma$. Figs. 4a through 4f show the results for the plasma frequencies $\omega = \omega' + i\omega''$ for $\gamma = 0.1, 0.3, 0.5, 0.7, 0.9, 1$ and $0 < |M| < 2$. It follows from Eq. (10) and from the numerical simulations that at given $\gamma$, the instability develops in the interval $\gamma < |M| < 1/\gamma$. The entire Mach number $|M|$ domain can be divided into three regions with distinctly different behavior of the plasmonic spectrum as shown in Fig. 4. At $|M| < 1$ and $\gamma|M| < 1$ (region I), the drift velocities in all cavities are smaller than the plasma velocity $v_p$ and the Dyakonov-Shur instability is developed in a very broad interval of the plasma frequencies yielding the frequency and increment ($\omega'' < 0$)/decrement ($\omega'' > 0$) patterns shown in Figs. 4a-4e for different values of $\gamma$. When $|M| > 1$ but $\gamma|M| < 1$ (region II), the drift velocity in the source and the drain cavities exceeds $v_p$, whereas in the central cavity it is smaller than $v_p$. In this case, the Dyakonov-Shur instability is supplemented by the plasmonic boom instability [12] resulting in the qualitatively different frequency and increment/decrement pattern as seen in Figs. 4a-4e. If both $|M| > 1$ and $\gamma|M| > 1$ (region III), the electron drift velocity exceeds $v_p$ in the entire FET structure and the instability disappears as illustrated in Figs. 4d, 4e. When the value of $\gamma$ is approaching unity the region of instability shrinks and the instability completely disappears at $\gamma = 1$ as shown in Fig. 4f. In this limit, the plasmon spectrum is determined by $\omega_n == \frac{\pi v_p}{3L}|1 - M^2|n$, $n = 1,2,...$ and describes drifting plasmons in the symmetric cavity of length $3L$ as expected.

The instabilities described above occur for both directions of the driving current ($M > 0$ and $M < 0$). As was mentioned in Ref. [20], this allows for driving this structure wirelessly via an RF signal with the frequency much smaller than the instability frequency. It also enables mixers, frequency multipliers, and heterodyne sources and receivers based on the variable width plasmonic structures. This approach is especially effective when the variable width plasmonic structures are assembled into plasmonic crystals considered in the next section.

## IV. PLASMONIC CRYSTAL TeraFET OF VARIABLE WIDTH

Plasmonic crystal FETs of variable width with various configurations of the elementary cell are shown in Fig.1d. Plasma dispersion in these plasmonic crystal structures is described by Eq. (9).

As shown in Ref. [12], the plasmonic crystal comprised of periodically repeated sections with different widths (see Fig. 1d) could support the plasmonic boom instability. This instability requires the electron drift velocity to exceed periodically the plasma velocity ($M > 1$) in some sections of the 2D electron channel. The condition $M > 1$ imposes additional restrictions on the design and material parameters of the plasmonic structure though very recently the plasmonic boom instability was measured in the plasmonic crystals with periodically modulated electron density [13].

We now use the hydrodynamic model to predict the types of the plasmonic crystals supporting the plasma instability at $M < 1$, such as the DS instability. The total power carried by the drifting plasmon $P = V_\omega^{eff} I_\omega^*$ consists of the electric ($P_{el} = V_\omega I_\omega^*$) and kinetic ($P_{kin} = V_\omega^{kin} I_\omega^*$) powers. If the total energy is conserved when the plasma wave crosses the boundaries between different sections of the 2D electron channel, the DS plasmon instability does not occur because the equal energy fluxes at both ends of the crystal elementary cell prevent the plasmons in the elementary cell from gaining energy (see Ref. [29] for more detailed discussions). This is in contrast with the plasmonic boom instability [12], which occurs via gaining energy from the electron flow crossing the plasmonic velocity threshold.

In order to support the DS instability, the plasmonic structure should break the total energy continuity of the drifting plasmons. In a single FET considered in Ref. [1], the asymmetric boundary conditions at the source and drain contacts ensured this discontinuity because of the different kinetic power flows at the source and the drain [29] resulting in the DS plasma instability at $M < 1$.

Another example is a plasmonic crystal consisting of periodically repeated gated and ungated sections in the 2D electron channel considered in Ref. [16]. The authors demonstrated that accounting for the ballistic transport at the boundaries between the gated and ungated sections results in the instability of the plasmonic band spectrum at $M < 1$. From the energy considerations, the "ballistic" boundary conditions (continuity of the voltage and the current) between the gated and ungated sections with different plasmon dispersion laws imply continuity of the electric power flow but discontinuity of the kinetic power flow. The resulting discontinuity of the total power flow may lead to the DS-type plasma instability (although only at certain phase matching conditions of the plasma waves in different elementary cells, i.e. in some finite intervals of the Bloch wave vector).

These examples demonstrate that the theory of the plasmonic crystal instability requires careful consideration of the boundary conditions between different sections of the 2D electron channel controlling the energy, voltages and currents at the interface that determine plasmon reflection and transmission.

We illustrate these requirements by calculating the plasmonic band spectrum for $M < 1$ in the plasmonic crystal with the arbitrary asymmetric elementary cell consisting of several sections of different widths where the total energy is conserved at the interfaces. The geometry of the crystal elementary cell and its electric equivalent circuit are shown in Figs. 5a and 5b, respectively. As shown in [12] and confirmed below, the instability for M<1 does not occur.

The plasmonic band spectrum is determined by Eq. (9) with matrix $\hat{T} = \prod_{i=1}^{4} \hat{t}_i$, where $\hat{t}_i$-matrices defined in Eq. (5) correspond to the four different sections of the electron channel in the crystal elementary cell, see Fig. 5b. The results of the plasmonic spectrum calculations are shown in Fig. 5c for several values of $M < 1$. As expected, the plasmonic band spectrum is purely real with a Doppler shift at finite drift velocities (see a detailed Doppler effect discussion in Ref. [29]).

In order to break the continuity of the energy flow in the crystal and provide conditions necessary for the onset of the DS instability we suggest using plasmonic stubs [19, 20] to define and control the boundaries between the plasmonic crystal cells. This approach allows for more design flexibility and optimization than simply relying on very abrupt boundaries to ensure the ballistic boundary conditions at the interfaces. (Accounting for the ballistic boundary conditions at the interface should predict the instability similar to that considered in the gate-ungated geometry in [16].)

A big advantage of using plasmonic stubs discussed in detail in Ref. [19] is an ability to control the boundary conditions by choosing the appropriate stub parameters. We now consider the plasmonic crystal structure with elementary cell comprised of the several sections of different width and the plasmonic stub. The geometry of the structure is shown in Fig. 6a along with its equivalent electric circuit shown in Fig. 6b. The stub in the plasmonic waveguide is a lumped element controlling the plasmon propagation [19]. In equivalent electric circuit diagram, the open-end stub is represented by the shunting impedance

$$Z_{st} = -\frac{i}{CW_s v_p} \cot \frac{\omega l}{v_p}, \tag{14}$$

where $W_s$ and $l$ are the width are the width and the length of the stub respectively [19, 30]. As shown in Ref. [20] the values of $V_\omega^{eff}$ and $I_\omega$ across the plasmonic stub in the electric circuit diagram in Fig. 6b are linked by the transfer matrix

$$\hat{s} = \begin{pmatrix} 1 + \frac{Z_0}{Z_{st}} \frac{M}{1-M^2} & -\frac{Z_0^2}{Z_{st}} \frac{M^2}{1-M^2} \\ \frac{1}{Z_{st}} \frac{1}{1-M^2} & 1 - \frac{Z_0}{Z_{st}} \frac{M}{1-M^2} \end{pmatrix} \tag{15}$$

This equation implies continuity of the electric voltage but discontinuity of the ac plasmonic current when the plasma wave in the channel travels across the stub. It happens because the ac current partially escapes into the gate through the stub. As a result, the total plasmonic energy flow in the channel $P = V_\omega^{eff} I_\omega^*$ has discontinuity when the plasma wave passes the stub. This provides conditions necessary (but not sufficient) for the DS instability. For example, our analysis shows that if the plasmonic crystal includes only stubs in its elementary cell the instability does not occur (see Appendix). The stub changes the energy conservation balance at the stub interfaces but fails to meet the necessary phase conditions for the plasmon reflection in a crystal, in contrast to a single FET with a stub considered in [20]. Both the energy change in the plasmonic elementary cell and plasmonic reflection with a proper phase matching are required for the standing plasma wave instability in a crystal. As shown below, adding a wider channel section to the elementary cell containing a stub enables the instability by providing the needed reflection conditions for the standing wave.

To this end, we calculated plasmonic band spectrum in the structure shown in Fig. 6 using Eq. (9) where $\hat{T}$-matrix is defined as $\hat{T} = \hat{t}_1 \hat{s} \hat{t}_2 \hat{t}_3$ with $\hat{t}_i$-matrices are given by Eq. (5). The $\hat{t}_i$-matrices describe the plasmon propagation in three different sections of the 2D channel in the crystal elementary cell as indicated in Fig. 6b. (In this calculation, we assumed $W_s = W$.)

The dispersion curves for several lowest plasmonic bands are shown in Fig. 7. These results confirm the possibility of the DS-type plasmonic instability in the variable width plasmonic crystal structure with stubs. Stable plasmonic bands ($Im\omega = 0$) at $M = 0$ are shown in Fig. 7a. When $M > 0$, the plasma frequencies become complex, see Figs. 7b and 7c. The modes with $Im\omega < 0$ are unstable with the instability increment mostly increasing at larger $M$. The instability increment is also larger for higher plasmonic bands. In contrast to the instability considered in [16], some plasmonic bands are unstable throughout the entire Brillouin zone and do not require additional phase matching between the plasma oscillations in different elementary cells. In some plasmonic bands $Im\omega > 0$, and these modes die out at $M > 0$. However, it follows from Eq. (9) that at the center of the Brillouin zone $Re\omega$ and $Im\omega$ are an even and odd functions of $M$, respectively. Therefore, the damped plasma modes at $M > 0$ become unstable at $M < 0$. This property is mostly preserved at the finite values of the Bloch vector as illustrated in Fig. 8 showing the plasmonic spectrum as a function $M$ at a fixed value of the Bloch vector $k \neq 0$. This result suggests that the plasma instability occurs for both directions of the DC current through the plasmonic crystal structure with the stub. As shown in [20] this property enables the RF to THz conversion.

Our analysis also shows that the plasmonic spectrum depends on the position of the stub in the elementary cell and the stub size providing ample opportunities for tuning the frequency and the increment of the plasma modes. In Fig. 9, the plasmonic band spectrum at fixed Mach number $M$ is plotted for two different positions of the stub in the elementary cell demonstrating the shift in both real and imaginary parts of the plasma frequency.

The predicted instability increments are on the order of the $v_p/L \sim \omega_p$. The losses introduced by the scattering of impurities, lattice vibrations, and viscosity of the electron fluid must be smaller than the increment in order not to suppress the instability. In the next Section, we show that this condition could be easily met at 77K for most typical semiconductor materials and even at 300 K for small enough dimensions. Even a more intense instability should occur in a graphene high mobility plasmonic crystal but this analysis is beyond the scope of this paper.

## V. NUMERICAL ESTIMATES FOR DIFFERENT MATERIALS

Since the instability increment is on the order of the fundamental plasma frequency, the necessary instability requirement reduces to the condition $Q \gg 1$. Here $Q = \omega_p \tau$ is the plasmonic quality factor and

$$\tau = \frac{\tau_m \tau_\nu}{\tau_m + \tau_\nu} \tag{16}$$

Here $\tau_m = \mu m^*/e$ is the momentum relaxation time, $\mu$ is the low field mobility, $\tau_\nu = \frac{1}{\nu q^2}$ is the viscosity relaxation time, $q = \frac{\pi}{2L}$ is the plasmonic wave vector for the fundamental plasma mode, $\nu = E_F \tau_{ee}/m^*$ is the viscosity of the electronic fluid defined in terms of the Fermi energy $E_F = \pi \hbar^2 n_o / m^*$ and the electron-electron collision time $\tau_{ee} = \hbar E_F / (3.4 k_B^2 T^2)$ [31]. Plasma frequency is determined as $\omega_p = v_p q$ where the plasma wave velocity $v_p$ is given by

$$v_p = \sqrt{\frac{\eta k_B T}{m^*}\left(1+e^{-\frac{eU_{gt}}{\eta k_B T}}\right) ln\left(1+e^{\frac{eU_{gt}}{\eta k_B T}}\right)} \qquad (17)$$

Here $U_{gt} = U_g - U_{th}$ is the gate voltage swing ($U_{th}$ is the threshold gate voltage), $k_B$ is the Boltzmann constant, and $\eta$ is the ideality factor. The last expression accounts for the dependence of the electron density in the channel on the temperature $T$ [32]. The quality factor first increases with the gate voltage swing along with the plasma frequency because the plasma oscillation period becomes smaller compared to the momentum relaxation time but then peaks in most cases because at high plasma frequencies the viscosity damping becomes dominant. Table I lists the maximum values of the quality factor $Q_m$ that could be achieved in Si, GaN, InGaAs, and p-diamond TeraFETs at 300 K and 77 K. These values were obtained by varying the gate voltage swing, see [33] for more details.) Table I also lists the plasma frequency $f_m$ at which the maximum value of Q is achieved. In longer devices, where viscosity is not dominant, we enter the values of $Q_m$ and $f_m$ achieved at 1 V gate voltage swing.

As seen, it is possible to realize plasmonic crystals at room temperature carefully choosing the operating regime to ensure the minimum damping. (A more detailed analysis of the quality factors is given in [33]). The estimated maximum quality factor in Si at room temperature for 22 nm and even 65 nm feature sizes might be sufficiently large to implement a plasmonic THz oscillator. p-diamond has the best combination of parameters for THz applications. It has promise of achieving sub-THz oscillations in the 300 GHz range, which is important for beyond 5G WIFI applications.

In this calculation, we used the highest reported values of the electron mobility for Si, GaN, and InGaAs and the hole mobility for p-diamond shown in Table II. Hence, the values presented in Table I should be considered as being close to the ultimate quality factors that could be obtained and as the figures of merit of the materials properties for the plasmonic crystal applications. The actual values of the field effect mobility might be lower because of the surface scattering and the boundary conditions imposing the ballistic mobility limitation [34, 35]. This gap between the predicted and measured values should be smaller in a plasmonic crystal, where the combination of the plasmonic stubs and varying width allows for a precise control of the plasmon propagation and reflection.

## VI.   CONCLUSIONS

The lack of efficient and compact electronic sources of sub-THz and THz radiation remains the main obstacle to a massive deployment of THz technology with many applications in industrial controls, medicine, biotechnology, homeland security, and communications. Plasmonic crystals have the best potential to solve this problem because they could combine nano sizes of individual cells with macro sizes of the entire crystal. This should lead to a dramatic power enhancement. Our analysis shows that in order to achieve both Dyakonov-Shur and plasmonic boom instability the plasmonic crystal design should use the variable width geometry combining the narrow protruding plasmonic regions – "plasmonic stubs" and wider longer sections. The stubs provide the proper boundary conditions between the cells enabling the instability by changing the plasma wave energy. The wider sections allow for the phase matched reflections from the boundaries between the sections of different widths with the plasmons gaining energy due to these reflections. The instability growth will be limited by nonlinearities leading to shock waves and solitons, in complete analogy with similar instabilities in water flows. The necessary condition for achieving the instability is having the plasmon quality factor exceeding unity. Our numerical estimates show that this could be achieved in Si, GaN, InGaAs, and p-diamond even at elevated temperatures. The new geometry proposed and analyzed in this paper enables the instability growth for both directions of the drive current. This makes it possible to drive the plasmonic crystal wirelessly by coupling an RF signal to the RF antennas coupled to the plasmonic crystal. This alleviates an interconnect problem that might otherwise affect the conversion efficiency. Also, using the separate tuning gates for the stub region allows for the frequency tuning and for frequency to digital conversion.

## VII. ACKNOWLEDGMENTS

The work was supported by ARO (Program Manager Dr. Joe Qiu). The work at RPI was also supported by the US Army Research Laboratory under ARL MSME Alliance (Project manager Dr. Meredith Reed).

## APPENDIX

In this Appendix, we illustrate our conclusion made in the main text that the plasmonic crystal with periodically repeated stubs in the channel of constant width does not support the plasma instability due to lack of the necessary asymmetric plasma wave reflections at the stub interfaces in the crystal structure.

We calculated plasmonic band spectrum for the plasmonic crystal structure with elementary cell shown in Fig. 10a which contains two stubs of different length. The equivalent electric circuit corresponding to this elementary cell is shown in Fig. 10b. Plasmonic spectrum of this crystal structure can be found from Eq. (9) where $\hat{T}$-matrix is defined as $\hat{T} = \hat{s}_1 \hat{t}_1 \hat{s}_2 \hat{t}_2$. Here, matrices $\hat{s}_{1,2}$ defined by Eq. (15) describe two different plasmonic stubs, and matrices $\hat{t}_{1,2}$ defined by Eq. (5) correspond to the two sections of the channel of different lengths as shown in Fig. 10a. For the sake of simplicity, in this calculation we assumed that the widths of the both stubs and that of the channel are the same.

Fig.10c shows the plasmonic band spectrum in the crystal Brillouin zone for several values of the Mach number $M$. This spectrum shows no instability ($\omega'' = 0$) and the Doppler shift at $M \neq 0$ in agreement with the prediction made in Ref. [29].

**FIGURE CAPTIONS**

FIG. 1. Different implementations of the variable width plasmonic devices (a -e) and schematics of the RF to THz conversion (f). TeraFETs shown in Fig. 1(b) to 1(c) are fabricated with varying width, $W_1$ and $W_2$. When the wider section length $\Delta L$ is small compared to the plasmonic wavelength $\lambda$ the protruding section is called plasmonic "stub". When $\Delta L$ is on the order of $\lambda$ the TeraFET is called "variable width" TeraFET.

FIG. 2. Variable width FET geometry (a) and the corresponding TL model (b).

FIG. 3. Real (a) and imaginary (b) parts of the plasma frequencies as a function of the Mach number $M$ in the plasmonic FET in Fig. 2 for non-interacting source/drain and central cavities ($\gamma = W/W_1 = 0$). Unstable plasma modes are indicated by the thick solid lines in Fig. 3b. Inset: spatial distribution of the normalized plasmonic current in the source cavity for several values of $M > 0$.

FIG. 4 Evolution of the plasmonic spectrum in the FET of the variable width with changing geometry (a) $\gamma = 0.1$; (b) $\gamma = 0.3$; (c) $\gamma = 0.5$; (d) $\gamma = 0.7$; (e) $\gamma = 0.9$; (f) $\gamma = 1$.

FIG. 5. Elementary cell of the plasmonic crystal consisting of several sections of different width (a) and its electric equivalent circuit diagram (b). (c) Plasmonic band spectrum at different values of the Mach number $M$. In this calculation $L_1 = 0.15L$, $L_2 = 0.1L$, $L_3 = 0.35L$, $L_4 = 0.4L$, $W/W_1 = 0.15$, and $W/W_2 = 0.35$.

FIG. 6. Elementary cell of the plasmonic crystal with the stub (a) and its electric equivalent circuit diagram (b).

FIG. 7. Frequencies ($\omega'$) and increments/decrements ($\omega''$) of the first three plasmonic bands in the plasmonic crystal structure shown in Fig. 6 for M=0, 0.3, and 0.7. In this calculation $L_1 = 0.1L$, $L_2 = 0.4L$, $L_3 = 0.5L$, $l = 0.2L$, and $W/W_1 = 0.1$

FIG. 8. Frequencies ($\omega'$) and increments/decrements ($\omega''$) of the first three plasma modes in the plasmonic crystal structure shown in Fig. 6 as a function of the Mach number $M$. In this calculation $kL = 1.7$ and all other parameters are the same as in Fig. 7.

FIG. 9. Frequencies ($\omega'$) and increments/decrements ($\omega''$) of the first three plasmonic bands in the plasmonic crystal structure shown in Fig. 6 for two different positions of the stub in the elementary cell and $M = 0.1$. Solid lines: $L_1 = 0.05L, L_2 = 0.45L$, dashed lines: $L_1 = 0.2L, L_2 = 0.3L$. All other parameters are the same as in Fig. 7.

FIG. 10. Elementary cell of the plasmonic stub crystal (a) and its electric equivalent circuit diagram (b). (c) Plasmonic band spectrum at different values of the Mach number $M$. The first three plasmonic bands are shown. In this calculation $L_1 = 0.2L, L_2 = 0.8L, l_1 = 0.63L$, and $l_2 = 021L$.

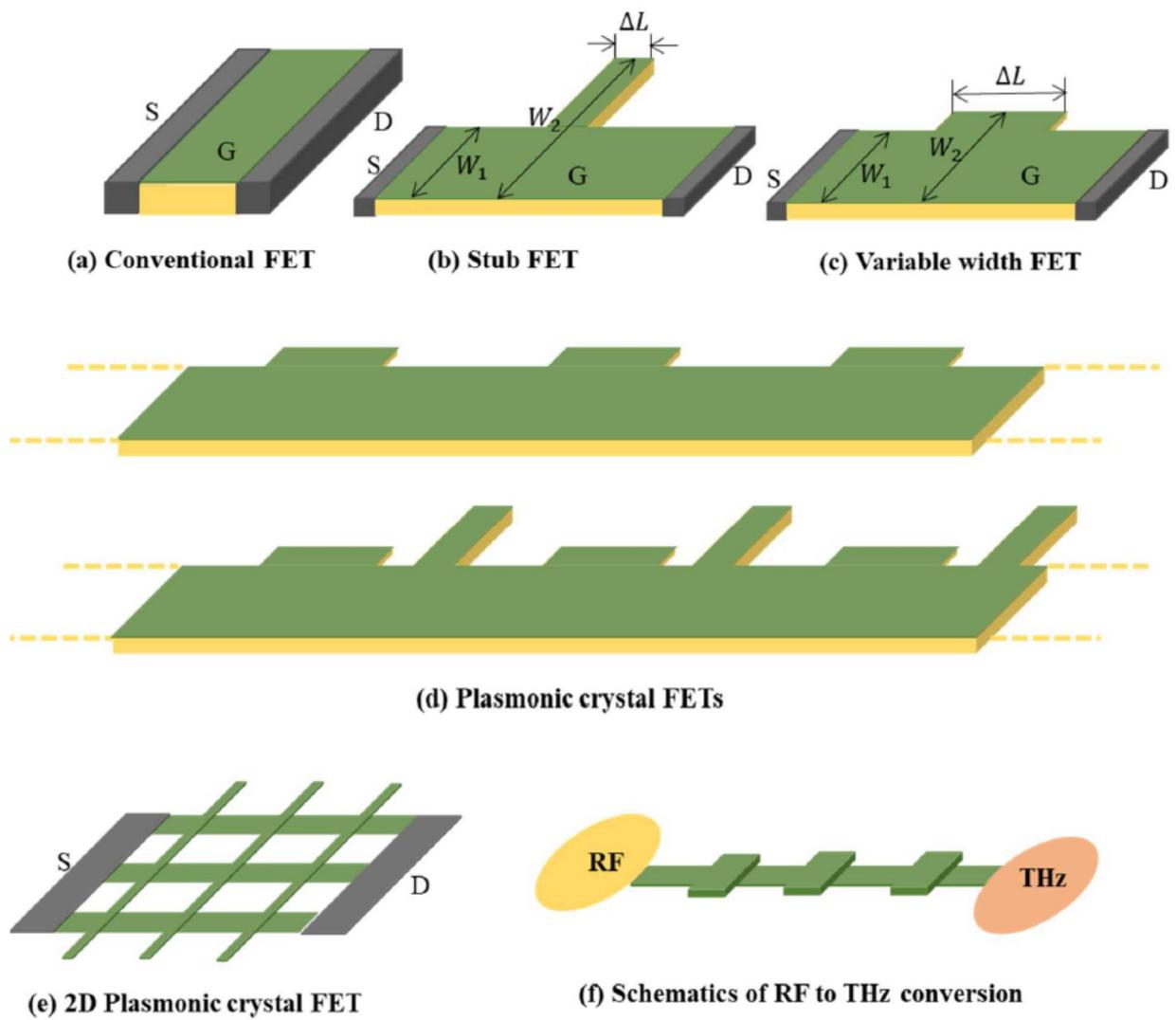

**Figure 1**

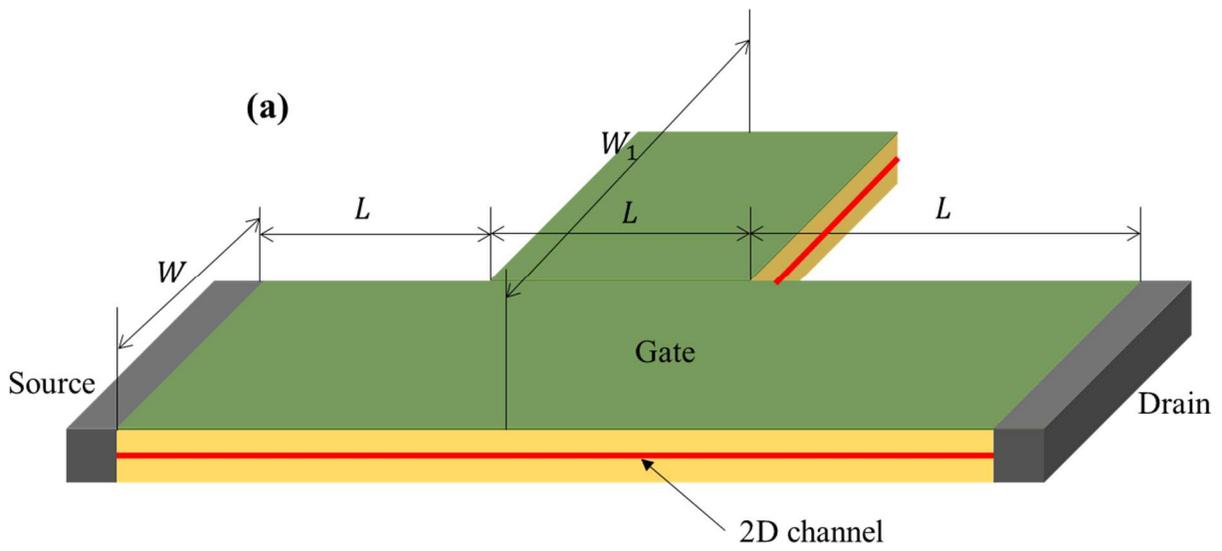

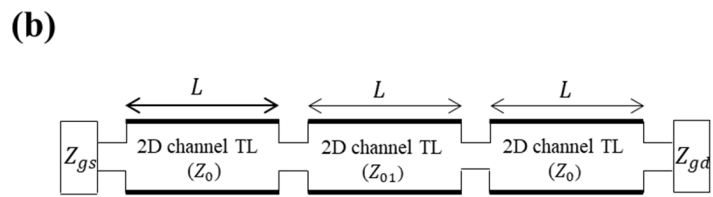

**Figure 2**

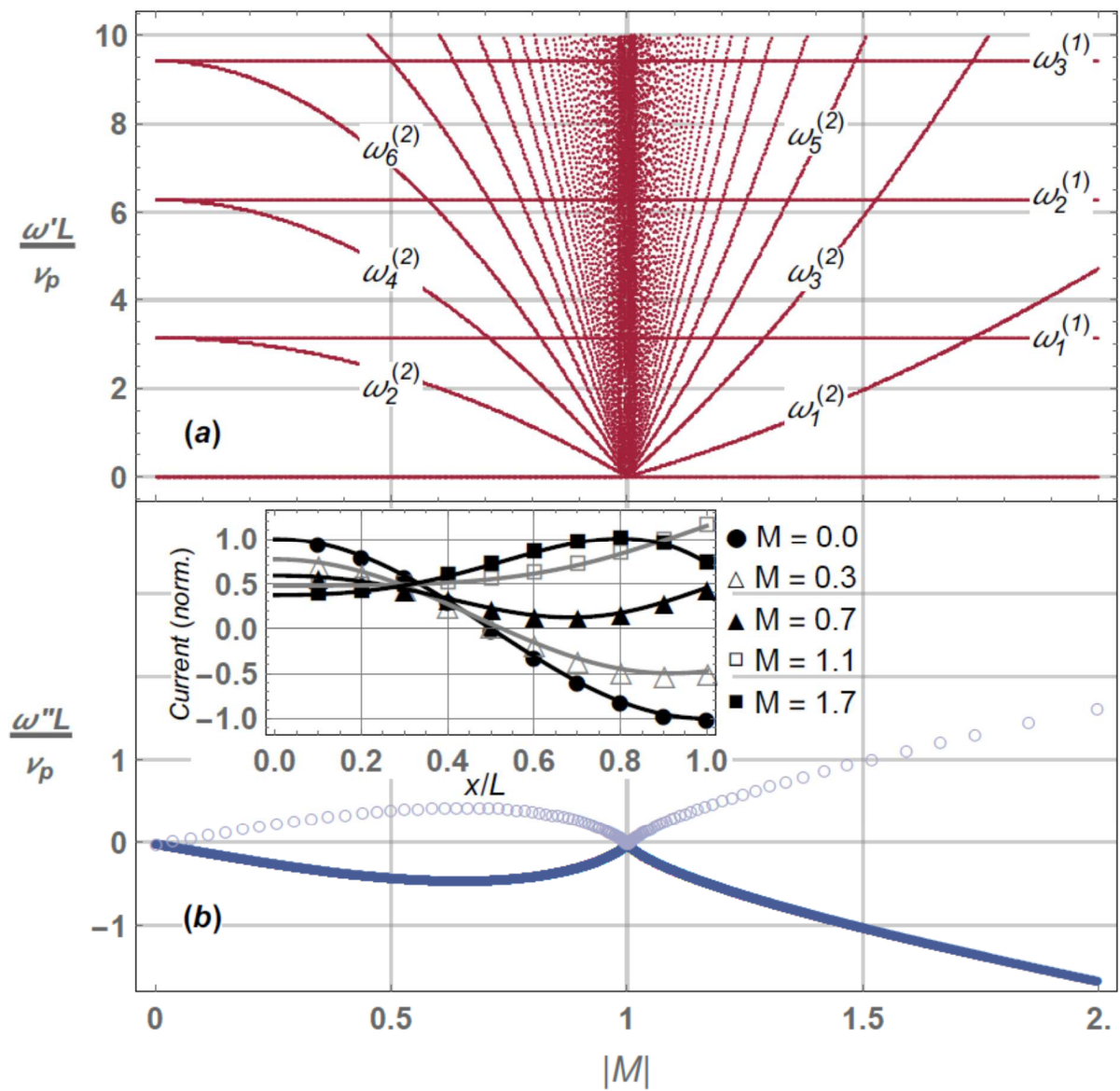

**Figure 3**

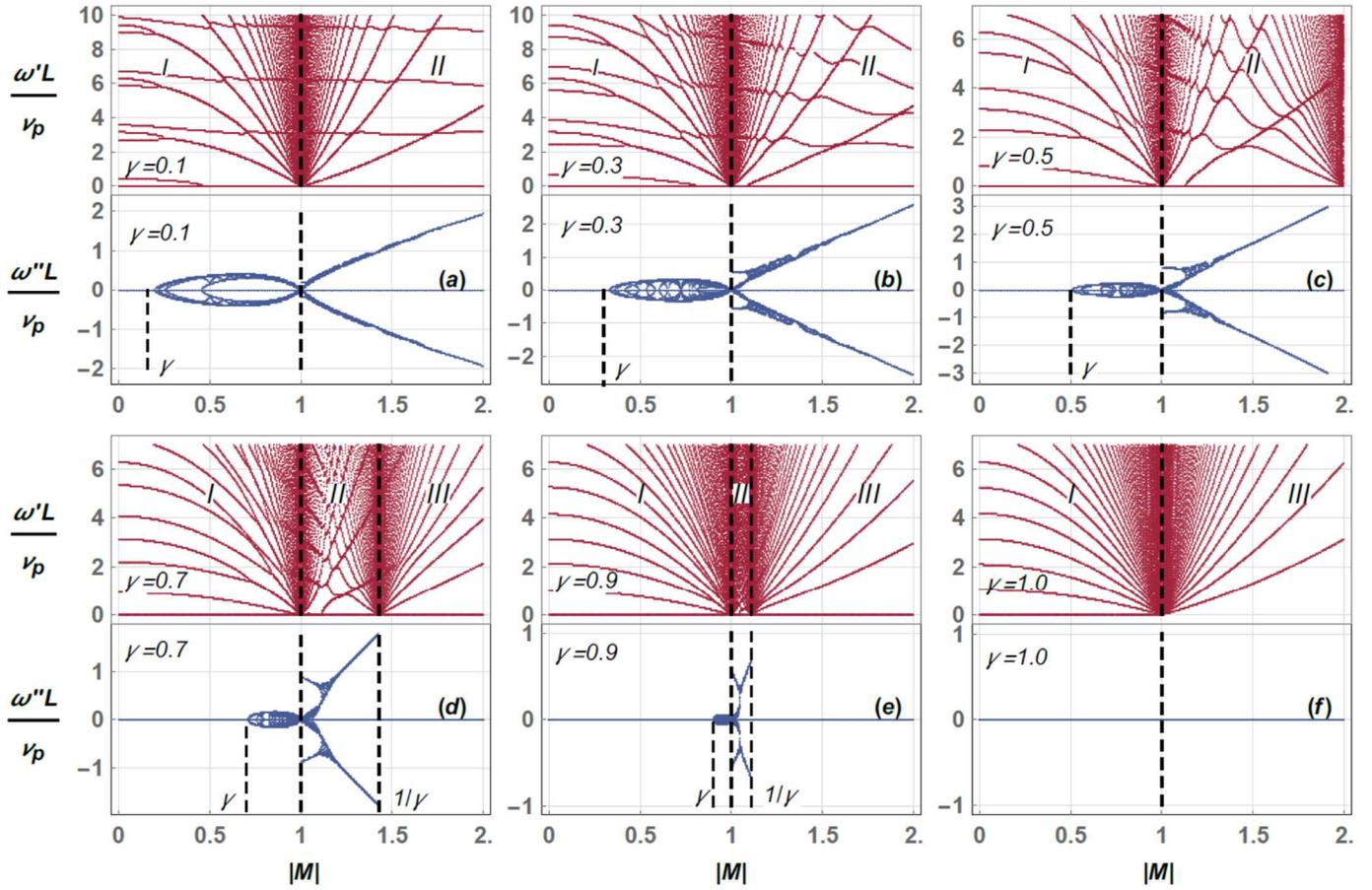

**Figure 4**

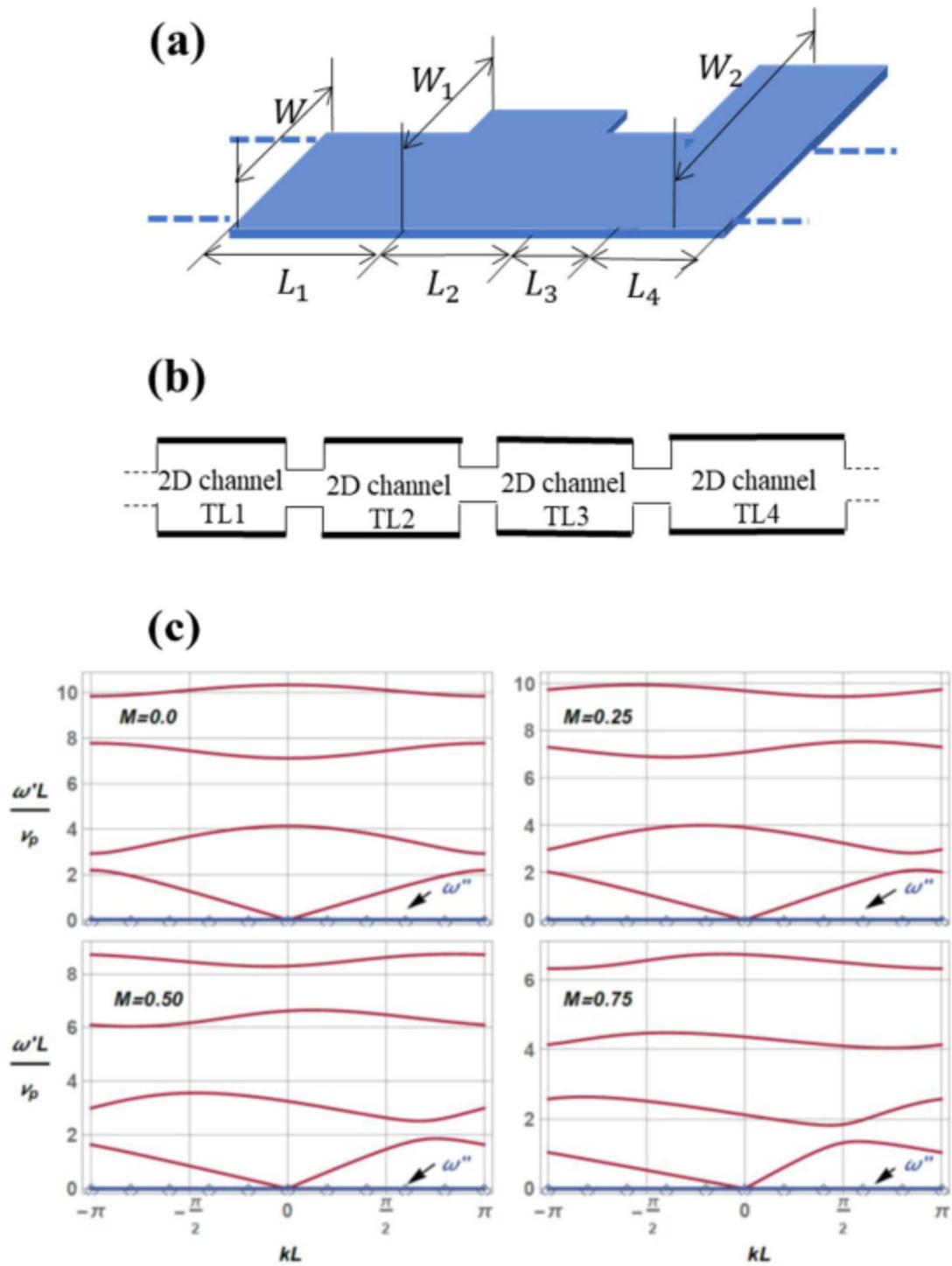

**Figure 5**

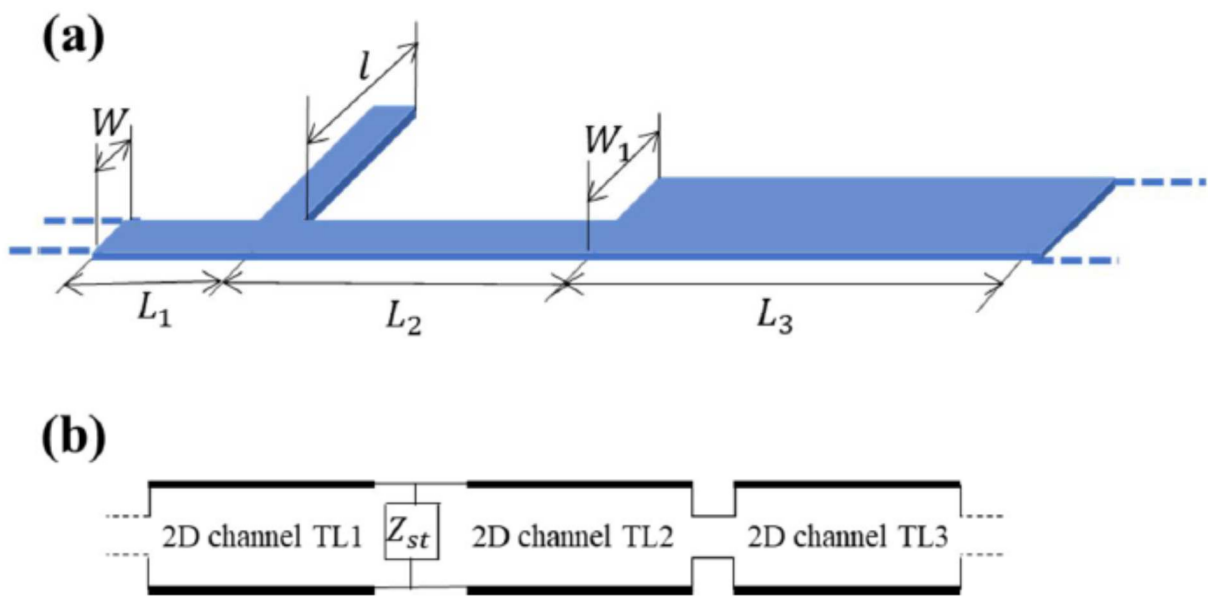

**Figure 6**

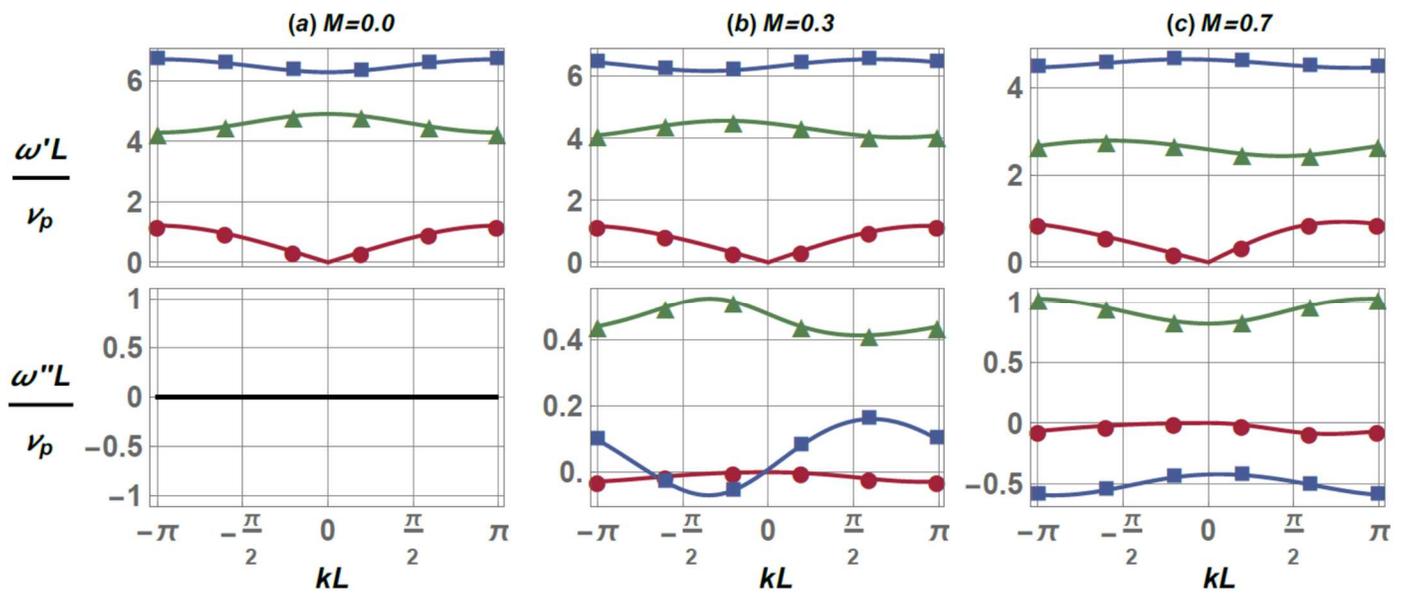

**Figure 7**

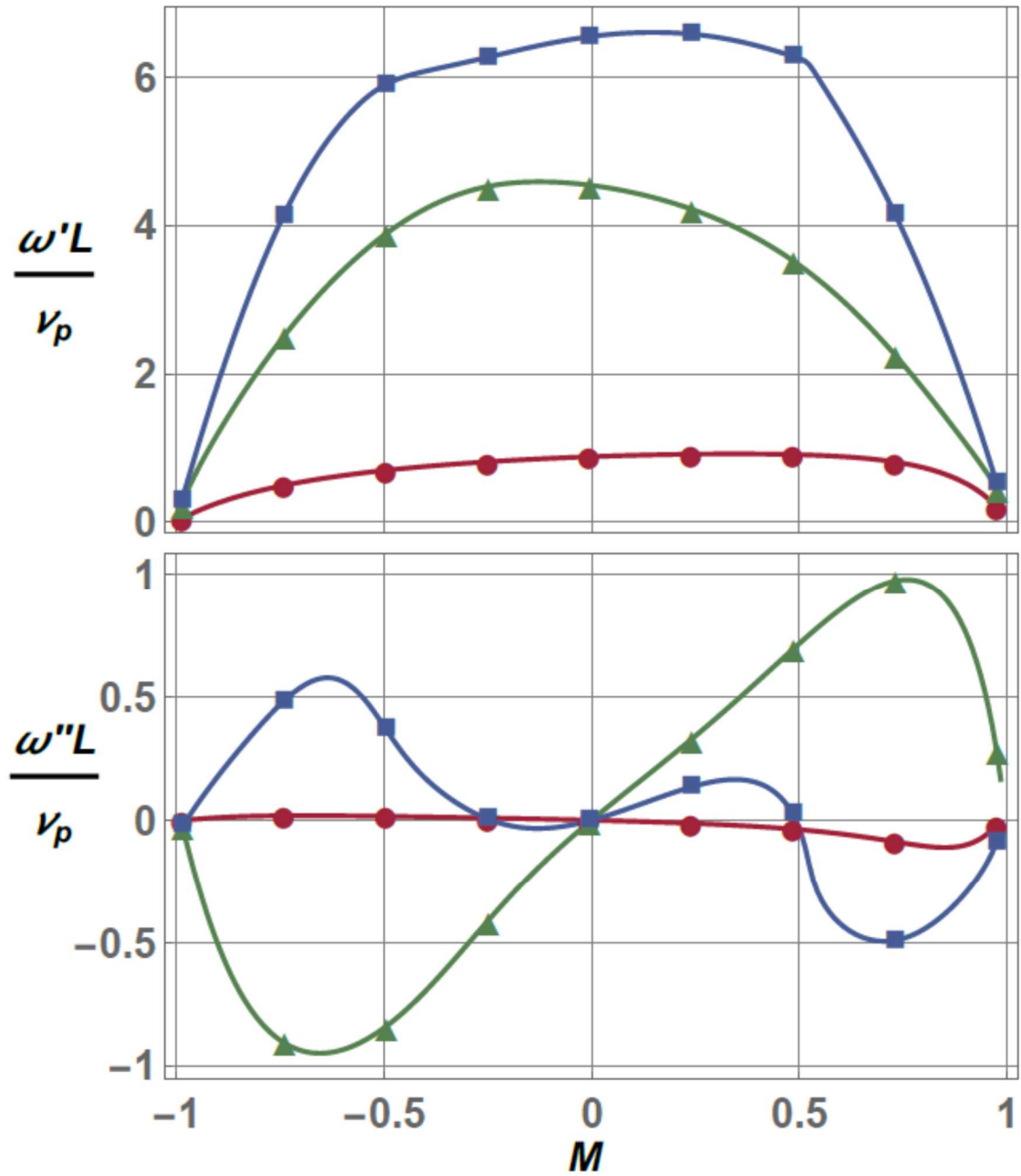

**Figure 8**

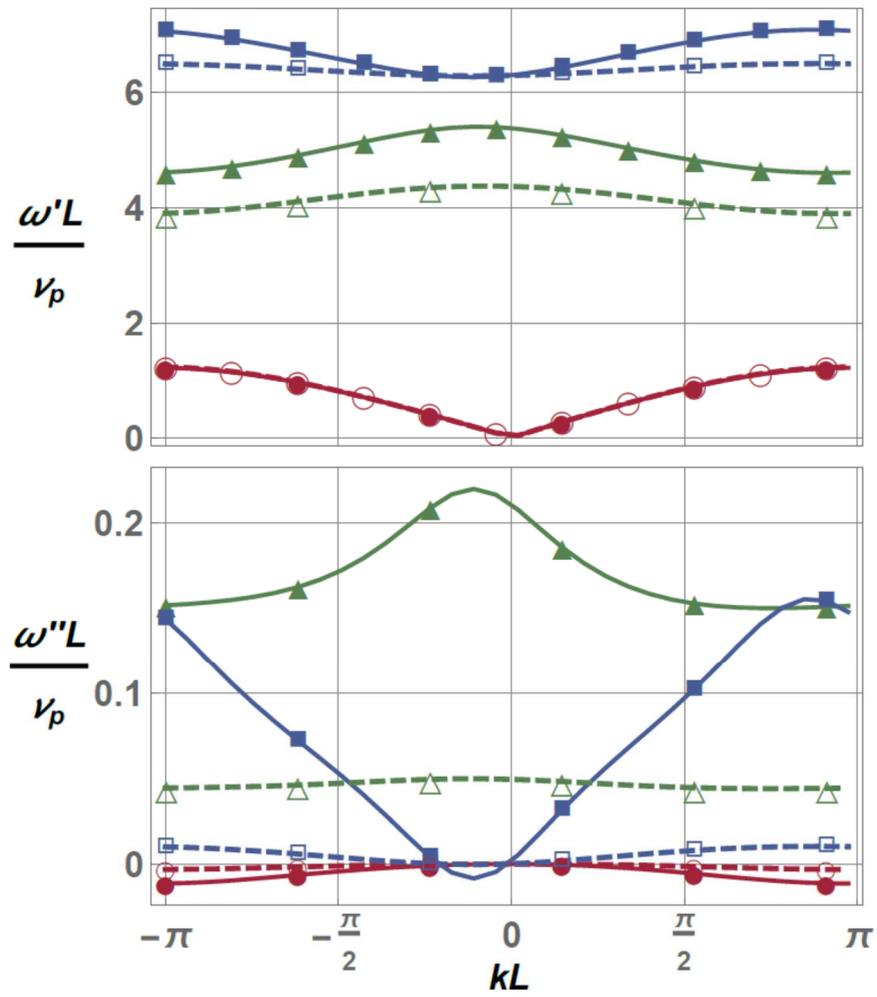

**Figure 9**

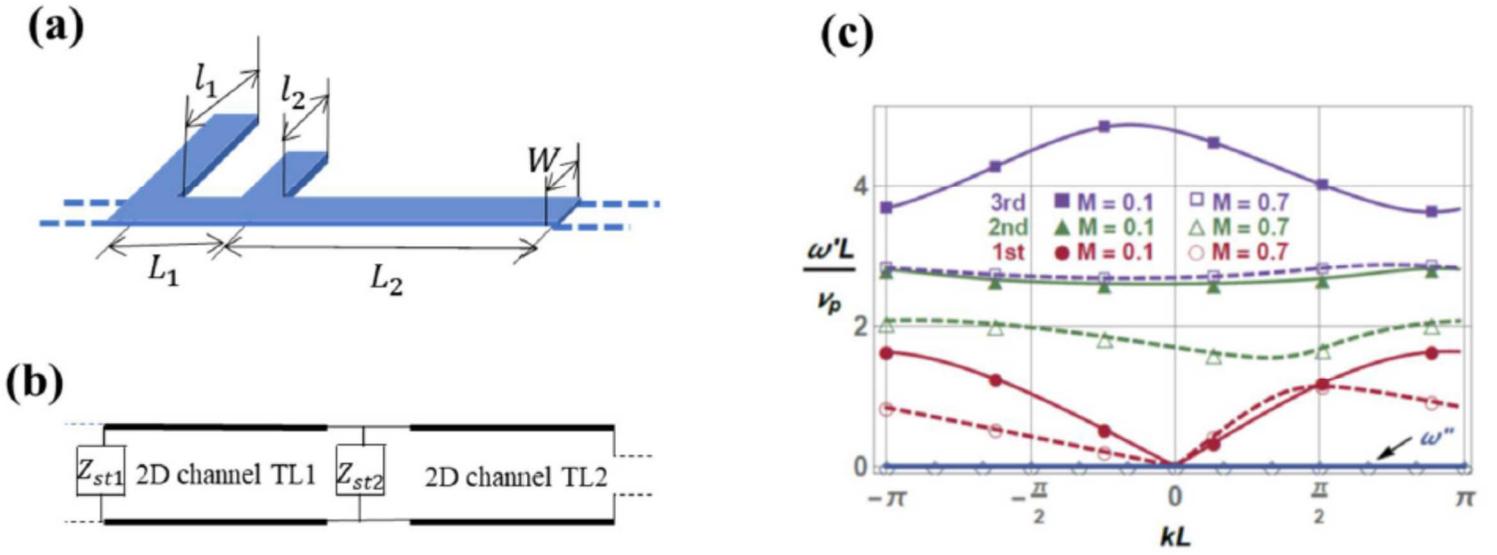

**Figure 10**

TABLE I. The values of $Q_m$ and $f_m$ for Si, GaN, InGaAs, and p-diamond TeraFETs at 300 and 77 K.

|  | 22 nm | | 65 nm | | 130 nm | |
| --- | --- | --- | --- | --- | --- | --- |
|  | $Q_m$ | $f_m$ (THz) | $Q_m$ | $f_m$ (THz) | $Q_m$ | $f_m$ (THz) |
| Si (300 K) | 7 | 12 | 3.5 | 3.6 | 1.75 | 1.8 |
| Si (77 K) | 27 | 3.5 | 15.2 | 2 | 11 | 1.65 |
| GaN (300 K) | 11 | 10 | 5.2 | 3.4 | 2.7 | 1.7 |
| GaN (77 K) | 47 | 2.5 | 28 | 1.5 | 18 | 1 |
| InGaAs (300 K) | 8 | 5 | 3.5 | 3 | 2.8 | 1.8 |
| InGaAs (77 K) | 18 | 2 | 7 | 1 | 4.5 | 0.8 |
| p-diamond (300 K) | 70 | 6 | 25 | 2 | 11 | 1 |
| p-diamond (77 K) | 180 | 3 | 100 | 1.5 | 80 | 1 |

TABLE II. Mobilities $\mu$ and effective electron masses $m^*$ used in the simulation at 300K and 77 K.

|  | Si | | GaN | | InGaAs | | p-diamond | |
| --- | --- | --- | --- | --- | --- | --- | --- | --- |
|  | 77 K | 300 K | 77 K | 300 K | 77 K | 300 K | 77 K | 300 K |
| $\mu$ (m$^2$/Vs) | 2 | 0.1450 | 3.1691 | 0.2 | 5.5 | 1.2 | 3.5 | 0.53 |
| $m^*$ | 0.19 | | 0.23 | | 0.041 | | 0.663 | |